\documentclass[aps,prd,tightenlines,twocolumn,floatfix,nofootinbib,showpacs,superscriptaddress]{revtex4}
\usepackage{dcolumn}% Align table columns on decimal point
\usepackage{graphicx}
\usepackage{epstopdf}

\newcommand{\be}{\begin{eqnarray}}
\newcommand{\ee}{\end{eqnarray}}
\newcommand\del{\partial}
\newcommand{\nn}{\nonumber}
\newcommand{\mui}{\mu_{\rm iso}}
\newcommand{\mat}{\left ( \begin{array}{cc}}
\newcommand{\emat}{\end{array} \right )}

\def\bml{\begin{mathletters}}
\def\eml{\end{mathletters}}
\def\ba{\begin{array}}
\def\ea{\end{array}}

\def\nn{\nonumber}
\def\to{\rightarrow}

\def\txi{\tilde{\xi}}

\def\ep{\epsilon}

\def\tr{\,{\rm tr}\,}
\newlength{\bredde}
\def\slash#1{\settowidth{\bredde}{$#1$}\ifmmode\,\raisebox{.15ex}{/}
\hspace*{-\bredde} #1\else$\,\raisebox{.15ex}{/}\hspace*{-\bredde} #1$\fi}
\def\Sl#1{\rlap{\raisebox{.15ex}{$\mskip 4 mu /$}}#1}  % improved slash
\begin{document}
%\draft
\title{
Extracting $F_\pi$ from small lattices: unquenched results}
\author{P. H. Damgaard}
\affiliation{
The Niels Bohr Institute, Blegdamsvej 17, DK-2100 Copenhagen \O,
Denmark
}
\author{U. M. Heller}
\affiliation{
American Physical Society, One Research Road, Box 9000, Ridge,
NY 11961-9000, USA
}
\author{K. Splittorff}
\affiliation{
The Niels Bohr Institute, Blegdamsvej 17, DK-2100 Copenhagen \O,
Denmark
}
\author{B. Svetitsky}
\affiliation{School of Physics and Astronomy, Raymond and Beverly Sackler
Faculty of Exact Sciences, Tel Aviv University, 69978 Tel Aviv, Israel
}
\author{D.~Toublan}
\affiliation {Physics Department, University of Maryland, College
Park, MD 20742
}

\date{\today}
\begin{abstract}
We calculate the response of the microscopic Dirac spectrum to an imaginary isospin
chemical potential for QCD with two dynamical flavors
in the chiral limit.
This extends our previous calculation from the quenched to the unquenched theory.
The resulting spectral correlation function in the $\epsilon$-regime provides here, too, a new and efficient
way to measure $F_\pi$ on the lattice. We test the method in a hybrid Monte
Carlo simulation of the theory with two staggered quarks. 
\end{abstract}
\maketitle

%\renewcommand{\thefootnote}{\fnsymbol{footnote}}
%\setcounter{footnote}{0}
%\narrowtext
%\widetext

\section{Introduction}

A primary challenge for lattice gauge theory is
the numerical determination of the low-energy constants of QCD with
nearly massless $u$ and $d$ quarks. 
The chiral limit is a notoriously difficult problem on the lattice.
Furthermore, one also has to face the issue of finite-volume
effects, a problem that is compounded by the light quarks. 
A situation like this is best tackled head-on by means of
finite-size scaling techniques, as is routinely done when the
correlation length becomes comparable to the system size (such as
near continuous phase transitions). Unfortunately, exact finite-size
scaling functions for the chiral limit are not easy to come by. One
notable exception is the so-called $\epsilon$-regime of QCD
\cite{LS}. It has been known for more than a decade \cite{Jac} that
this makes possible precise measurements of the
infinite-volume chiral condensate
$\Sigma=\langle\bar\psi\psi\rangle$ from simulations on small
lattices. By definition, in the $\epsilon$-regime the linear extent
$L$ of the box is smaller than the Compton wavelength
$1/m_{\pi}$ of the pions, 
but much larger than $1/\Lambda_{\rm QCD}$. 
It is in this sense that the lattice required is small: Rather than
demanding that the pion be contained by the lattice, one requires
precisely that the lattice be too small for the pion.

In the $\epsilon$-regime the leading term of the 
effective partition function is a known function that is
determined uniquely by the flavor symmetries and their spontaneous
breaking. While this leads to an equally computable analytical
expression for the condensate $\langle\bar{\psi}\psi\rangle$ that can be
compared to lattice measurements 
\cite{Jac1}, a far more precise estimate comes from extracting $\Sigma$ from
the analytical expressions for the distributions and correlations of the
smallest eigenvalues of the Dirac operator. 
The universal predictions for different eigenvalue 
distributions according to the topological index $\nu$ and gauge group
representation of the quarks \cite{DN} have been checked in detail
for both staggered and overlap fermions in a variety of studies
with different color representations, mainly in quenched calculations
(see, e.~g., Refs.~\cite{BB,Jac} for a partial list), but also with
dynamical fermions \cite{BB1,DHNR}.

Here we will consider the computation of the low-energy constant 
$F_\pi$, associated with the decay of pions.  
In the conventional approach $F_\pi$ is measured in a regime where 
the Compton wavelength of the pion is much smaller than the lattice size 
(see \cite{Fpi-quen} for some recent quenched and \cite{Fpi-dyn} for
some recent dynamical calculations of $F_\pi$). We have
recently presented a new method for extracting the pion decay
constant $F_\pi$ from Dirac operator eigenvalues in the
$\epsilon$-regime \cite{DHSS}. In our first paper the focus was on
quenched QCD. In this article, 
we extend this method to QCD with two dynamical light flavors, the
physical $u$ and $d$ quarks. 

In principle, all low-energy constants of QCD can be determined
numerically from the spectrum of the Dirac operator. 
To leading order the spectrum depends only on the infinite-volume
chiral condensate $\Sigma$, but higher order terms (in the counting
of the $\epsilon$-expansion for the chiral Lagrangian \cite{LS})
give rise to measurable deviations. The idea of the improved method
proposed in Ref.~\cite{DHSS} is to make use of an enlarged set of
couplings provided by external fields. In particular we introduce a
constant Abelian field that couples with opposite sign to the $u$
and $d$ quarks. This can be interpreted as an imaginary isospin
chemical potential \cite{MT}. Such a source turns out to have a
dramatic effect on a certain spectral correlation function of the
Dirac operator. It is this effect which allows for a very precise
determination of $F_{\pi}$.

An alternative method is to use a real baryon chemical
potential to extract the $F_{\pi}$-dependence of the microscopic
spectral density. 
Recently quenched results based on this method
have been reported \cite{OW}. 
The analytical predictions for this method were derived in
Ref.~\cite{baryon,SplitVerb2}. 
The advantage of our method is that the Dirac operator eigenvalues remain
real, while a baryon chemical potential makes them (and the determinant)
complex. The pion decay constant has also been measured
from meson correlation functions in the $\epsilon$-regime (see $e.g.$
\cite{Fpi-ep}).

The outline of the paper is as follows. We first present the analytical result
for the spectral correlation
function whose dependence on $F_\pi$ is the backbone of our method. We then
compare this prediction to the results of numerical simulations of QCD with
two dynamical flavors, and use this to extract the value for $F_\pi$.

\section{Two Light Flavors: The Analytical Result}

An imaginary isospin chemical potential leads us to consider two distinct
Dirac operators and their eigenvalues for a given gauge
potential $A$:
\be
D_+\psi_{+}^{(n)} &\equiv&
(\Sl{D}(A)+i\mui\gamma_0)\psi_{+}^{(n)}=i\lambda_{+}^{(n)}\psi_{+}^{(n)},\\[3pt]
D_-\psi_{-}^{(n)} &\equiv& (\Sl{D}(A)-i\mui\gamma_0)\psi_{-}^{(n)}
= i\lambda_{-}^{(n)}\psi_{-}^{(n)}.
\ee
Since the operators $D_{\pm}$ are anti-hermitian, the eigenvalues
$\lambda_{+}^{(n)} $ and $\lambda_{-}^{(n)} $ lie on the real axis. This justifies the
otherwise artificial use of an {\em imaginary} isospin chemical
potential. Viewed in terms of external gauge potentials it is actually
the natural choice, corresponding to a constant real Abelian gauge potential
$A_0=\mui$.

We consider the mixed two-point spectral correlation function of the Dirac operators $D_{\pm}$.
The formulae for the quenched theory were given in Ref.~\cite{DHSS}; here we
briefly present the corresponding expressions for
two dynamical flavors.
The mixed two-point function, which depends on the two light masses $m_u$ and
$m_d$ and on $\mui$, is defined as
\begin{widetext}
\be
\rho^{(2)}(\lambda_+,\lambda_-,m_u,m_d;i\mui) \equiv \Bigl\langle \sum_n \delta(\lambda_+-\lambda_+^{(n)})\sum_l
 \delta(\lambda_--\lambda_-^{(l)})\Bigr\rangle %\nn \\&&
   - \Bigl\langle \sum_n \delta(\lambda_+-\lambda_+^{(n)}) \Bigr\rangle
         \Bigl\langle\sum_l \delta(\lambda_--\lambda_-^{(l)}) \Bigr\rangle,
\label{corr-def}
\ee
where the brackets 
denote the average over the QCD partition function with two flavors
and $\mui\neq0$,
 \be { Z}_{2}(m_u,m_d;i\mui)& = & \int[{\rm d}A]_\nu
\det(D_++m_u)\det(D_-+m_d)\,e^{-S_{\rm YM}(A)}. \ee
The index $\nu$
on the measure indicates that this is the partition function in a
sector where the gauge fields have topological charge $\nu$. Keeping
fixed the scaling variables
\be \xi_\pm=\lambda_\pm\Sigma V,\quad \hat
m_{u,d}=m_{u,d}\Sigma V,\quad \hat\mui=\mui F_\pi \sqrt{V}, \ee
as $V\to\infty$, we define the microscopic two point
correlation function 
\be \rho^{(2)}_s(\xi_+,\xi_-,\hat
m_u,\hat m_d;i\hat\mui) \equiv \lim \frac{1}{\Sigma^2  V^2} \,
\rho^{(2)}\Big(\frac{\xi_+}{\Sigma V},\frac{\xi_-}{\Sigma V},\frac{\hat
  m_u}{\Sigma V},\frac{\hat m_d}{\Sigma V};i\frac{\hat\mui}{F_\pi\sqrt{V}}\Big). \label{rho2micro}
\ee
This spectral correlation function $\rho^{(2)}$ can be 
obtained from the mixed  ``partially quenched'' scalar susceptibility, \be
\chi(m_+,m_-,m_u,m_d;i\mui) \equiv \left\langle \tr
\frac{1}{D_++m_+}\tr
 \frac{1}{D_-+m_-}\right\rangle%  \nn\\[2pt]&  &
  - \left\langle \tr \frac{1}{D_++m_+}\right\rangle
         \left\langle\tr\frac{1}{D_-+m_-}\right\rangle,
\label{chimix}
\ee
which possesses the spectral representation
\be
\chi(m_+,m_-,m_u,m_d;i\mui)  =  \left\langle \sum_n \frac{1}{i\lambda_+^{(n)}+m_+}\sum_l
 \frac{1}{i\lambda_-^{(l)}+m_-}\right\rangle  %\nn\\
 %% &  &
  - \left\langle \sum_n \frac{1}{i\lambda_+^{(n)}+m_+}\right\rangle
         \left\langle\sum_l\frac{1}{i\lambda_-^{(l)}+m_-}\right\rangle.
\ee
The desired correlation function follows from the double discontinuity across
the imaginary axis \cite{DOTV},
\be
\rho^{(2)}(\lambda_+,\lambda_-,m_u,m_d;i\mui) & = & \frac1{4\pi^2}
{\rm  Disc}\,\chi(m_+,m_-,m_u,m_d;i\mui)|_{m_+=i\lambda_+\atop m_-=i\lambda_-} \label{rho2disc}\\[3pt]
& = & \frac1{4\pi^2}\lim_{\ep\to0^+}[
 \chi(i\lambda_++\ep,i\lambda_-+\ep,m_u,m_d;i\mui)-\chi(i\lambda_+-\ep,i\lambda_-+\ep,m_u,m_d;i\mui) \nn\\
&&\qquad -\chi(i\lambda_++\ep,i\lambda_--\ep,m_u,m_d;i\mui)
+\chi(i\lambda_+-\ep,i\lambda_--\ep,m_u,m_d;i\mui)]. \ee
\end{widetext}
The power of the $\epsilon$-regime is that this function, $\rho^{(2)}$, can
be expressed in closed analytical form.

The mixed scalar susceptibility (\ref{chimix}) involves two additional
fermion species, with mass $m_+$ and $m_-$, coupled as well to the external field
$\mui$. This quantity thus has to be evaluated in a theory with {\em
four} fermion species, of which only two become physical (the $u$
and $d$ quarks), while the remaining two are quenched away 
and serve as sources for the two eigenvalues involved in the
two-point correlation function. 
In order to do this we will use the replica method (see, e.g.,
\cite{DS}). We have
\begin{widetext}
\be\label{chi-replica}
\chi(m_+,m_-,m_u,m_d;i\mui)& = &
\lim_{n\to0} \frac{1}{n^2}\del_{m_+}\del_{m_-}\log
Z_{2n,2}(m_+,m_-,m_u,m_d;i\mui) ,
\ee
where the generating functionals $Z_{2n,2}$ have $2n$ replica flavors
besides the two flavors of mass $m_u$ and $m_d$,
\be
{ Z}_{2n,2}(m_+,m_-,m_u,m_d;i\mui) &=&  \int[{\rm d}A]_\nu
\left[\det(D_++m_+)\det(D_-+m_-)\right]^n \nn\\&& \quad\times
 \det(D_++m_u)\det(D_-+m_d)
      \,e^{-S_{\rm YM}(A)}.\label{ZnQCD}
\ee
Half of the replica flavors have mass $m_+$ and chemical potential
$+i\mui$ while the other half have mass $m_-$ and chemical potential
$-i\mui$.

In the $\epsilon$-regime the
leading terms of the effective partition functions satisfy a string of Toda lattice
equations which can be used to extract the replica limit
\cite{K,SplitVerb1}. Specifically \cite{SplitVerb2},
\be
\hat m_+\del_{\hat m_+} \hat m_-\del_{\hat m_-}\log Z_{2n,2}(\hat m_+,\hat m_-,\hat m_u,\hat m_d;i\hat \mui)&&\nn\\[3pt]
&& \hspace{-5cm}
= 4 n^2
%V^4\Sigma^4
(\hat m_+\hat m_-)^2\frac
  {Z_{2n+2,2}(\hat m_+,\hat m_-,\hat m_u,\hat m_d;i\hat \mui)
Z_{2n-2,2}(\hat m_+,\hat m_-,\hat m_u,\hat m_d;i\hat \mui)}{[Z_{2n,2}(\hat m_+,\hat m_-,\hat m_u,\hat m_d;i\hat \mui)]^2}.
\ee
Taking the replica limit (\ref{chi-replica}) of this Toda lattice equation we get
\be
\chi(\hat m_+,\hat m_-,\hat m_u,\hat m_d;i\hat \mui)%&&\nn\\[3pt]&&\hspace{-2cm}
=
4
%V^4\Sigma^4
\hat m_+\hat m_- \frac{Z_{2,2}(\hat m_+,\hat m_-,\hat m_u,\hat m_d;i\hat \mui)Z_{2,-2}(\hat m_u,\hat m_d|\hat m_+,\hat m_-;i\hat \mui)}{[Z_2(\hat m_u,\hat m_d;i\hat \mui)]^2}.
\label{chi-dyn}
\ee
We need three different effective partition functions to evaluate the right
hand side of this equation. Of these, the two that involve only positive numbers of
flavors are of the form
\be
Z_{2n,2}(\hat m_+,\hat m_-,\hat m_u,\hat m_d;i\hat \mui) =
\int_{U \in U(2n+2)} dU \det(U)^\nu e^{\frac{1}{4}VF_\pi^2\mui^2
{\rm Tr} [U,B][U^\dagger,B] + \frac 12 \Sigma V {\rm Tr}(M^\dagger U + MU^\dagger)}\,,
\label{zeff}
\ee
where $B={\rm diag}(1_{n+1},-1_{n+1})$ and 
$M = {\rm diag}(m_+,\ldots,m_+,m_u,m_-,\ldots,m_-,m_d)$. In
particular, 
\be Z_2(\hat m_+,\hat m_-;i\hat \mui)
%&=&  e^{-2VF^2_\pi\mui^2}\int_0^1 dt\, t
%e^{2VF^2_\pi\mui^2t^2} I_\nu(t m_+\Sigma V)I_\nu (t m_-\Sigma V)\nn\\
&=& e^{-2\hat\mui^2}\int_0^1 dt\, t
e^{2\hat\mui^2t^2} I_\nu(t \hat m_+)I_\nu (t \hat m_-),
\label{Z1mui}
\ee
where we have defined $\hat m_\pm=m_\pm\Sigma V$,
while an explicit analytical expression for $Z_{2,2}$ can be obtained from
Ref.~\cite{AFV},
\be
Z_{2,2}(\hat m_+,\hat m_-,\hat m_u,\hat m_d;i\hat \mui) =
\frac{1}{(\hat m_+^2-\hat m_u^2)(\hat m_-^2-\hat m_d^2)}%\nn\\[3pt]&&\quad\times
\,\left|
\begin{array}{cc}
Z_2(\hat m_+,\hat m_-;i\hat \mui) &  Z_2(\hat m_+,\hat m_d;i\hat \mui)\\[2pt]
Z_2(\hat m_u,\hat m_-;i\hat \mui) &  Z_2(\hat m_u,\hat m_d;i\hat \mui)
\end{array}\right|.
\ee

The remaining partition function $Z_{2,-2}$ on the right hand side of Eq.~(\ref{chi-dyn}) is the
most difficult part. It involves two fermion species (the two physical $u$ and $d$ quarks) and in
addition $-2$ flavors, i.e., two bosons. The resulting effective partition function is an integral over a graded group.
Deferring the technical
discussion of this to a forthcoming paper \cite{nextpaper}, we quote the final result,
\be \label{Z2m2}
Z_{2,-2}(\hat m_u,\hat m_d|\hat m_+,\hat m_-;i\hat \mui) =
\left|\begin{array}{cc}
(\hat m_u^2-\hat m_+^2)Z_2(\hat m_u,\hat m_d;i\hat \mui) &  Z_{1,-1}(\hat m_d|\hat m_-)\\[2pt]
-Z_{1,-1}(\hat m_u|\hat m_+) &  (\hat m_d^2-\hat m_-^2)Z_{-2}(\hat m_+,\hat m_-;i\hat \mui)
\end{array}\right|.
\ee
In Eq.~(\ref{Z2m2}) the ($\mu$-independent) partition function with one
fermion and one boson is given by
\be
Z_{1,-1}(\hat x|\hat y) =  \hat x I_1(\hat x) K_0(\hat y) +\hat y
K_1(\hat y)I_0(\hat x),
\label{Zsusy}
\ee
and the partition function with two bosons is
\be
Z_{-2}(\hat x,\hat y;i\hat \mui) = e^{2\hat{\mui}^2}\int_1^\infty dt\, t
e^{-2\hat\mui^2t^2} K_\nu(t \hat x)K_\nu (t \hat y).
\label{Z-1mui}
\ee
where $\hat x=x\Sigma V$ and $\hat y=y\Sigma V$.

We can finally compute the required two-point correlation function (\ref{rho2micro}) by
means of the discontinuity (\ref{rho2disc}).
The result is
\be\label{rho-result}
\rho^{(2)}_s(\xi_+,\xi_-,\hat m_u,\hat m_d;i\hat\mui)
& = & \xi_+\xi_- \left[
\int_0^1 dt\, t e^{2\hat\mui^2 t^2} I_\nu(t\hat m_u)
I_\nu (t\hat m_d)\right]^{-2} \\
& & \times \left[
\int_0^1 dt\, t e^{2\hat\mui^2 t^2} J_\nu(t \xi_+)
J_\nu (t \xi_-) \int_0^1 dt\, t e^{2\hat\mui^2 t^2}
I_\nu(t\hat m_u)I_\nu (t\hat m_d) \right.\nn\\
& & \qquad-\left.
\int_0^1 dt\, t e^{2\hat\mui^2 t^2} I_\nu(t\hat m_u)
J_\nu (t \xi_-) \int_0^1 dt\, t
e^{2\hat\mui^2t^2} J_\nu (t \xi_+) I_\nu(t\hat m_d) \right]\nn\\
&& \times\left[ \int_0^1 dt\, t e^{2\hat\mui^2t^2} I_\nu(t\hat m_u)
I_\nu (t\hat m_d)\int_1^\infty dt\, t e^{-2\hat\mui^2 t^2}
J_\nu(t \xi_+)J_\nu (t \xi_-)\right. \nn\\
&&\hspace{-2cm}\left.+\frac{[\hat m_u I_{\nu+1}(\hat m_u)
J_\nu(\xi_+)+\xi_+ J_{\nu+1}(\xi_+)I_\nu(\hat m_u) ]\,
[\hat m_d I_{\nu+1}(\hat m_d) J_\nu(\xi_-)+\xi_- J_{\nu+1}(\xi_-)I_\nu(\hat m_d) ]}{(\xi_+^2+\hat m_u^2)(\xi_-^2+\hat m_d^2)}\right].\nn
\ee
For a numerical evaluation it is advantageous to rewrite the improper
integral appearing in the fourth line as
\be
 \int_1^\infty dt\, t e^{-2\hat\mui^2t^2}
J_\nu(t \xi_+)J_\nu (t \xi_-)
  =
\frac{1}{4\hat\mui^2}
\exp\left(-\frac{\xi_+^2+\xi_-^2}{8 \hat\mui^2}\right)
I_\nu\left(\frac{\xi_+\xi_-}{4\hat\mui^2}\right)
-\int_0^1 dt\, t e^{-2\hat\mui^2 t^2}
J_\nu(t \xi_+)J_\nu (t \xi_-).
\ee
\end{widetext}
In exact analogy to the quenched case \cite{DHSS}, the correlation 
function (\ref{rho-result}) changes dramatically when $\mui$ is made non-zero.
A fit to Monte Carlo data with $\mui\not=0$ using
Eq.~(\ref{rho-result})
will then readily produce a measurement of $F_\pi$.

\section{Numerical Results}

We demonstrate our method by applying it to the eigenvalue correlation
function calculated for SU(3) lattice gauge theory with two staggered
fermions of equal mass $m_u=m_d\equiv m$. 
The gauge action is the standard plaquette action and unimproved staggered
operators are used throughout. 
This theory has an exact U(2)$\times$U(2) symmetry; while it represents eight
tastes in the continuum limit, we work far from the continuum limit, at a
strong coupling $\beta=4.2$, in order to stay on the confining side of bulk
and finite-temperature phase transitions \cite{8flavors}. 
In addition, and since we only aim at illustrating the method here, this
rather strong coupling provides us with an unambiguous number of 
pseudo-Goldstone modes.  
At the coupling considered, two staggered fermions give rise to  
spontaneous chiral symmetry breaking of the form U(2)$\times$U(2)$\to$U(2).
The staggered theory at this coupling thus appears to be ``insensitive''
to gauge field topology (see for instance \cite{DHNR_topo}): The
additional U(1) factor in the coset dictates that we must compare 
our numerical results with $\nu = 0$ in Eq.~(\ref{rho-result}).

We ran the exact Hybrid Monte Carlo algorithm \cite{HMC} on lattices with
volume $V=4^4$ and~$6^4$, setting $\mui=0.0125$ and~0.0055, respectively, in
order to keep constant the scaling variable 
$\hat\mui$.  We studied the cases $m=0.01$ and~0.025 for the smaller volume,
and correspondingly $m=0.002$ and~0.005 in the larger volume. Note that the
scaling variable $\hat m=m\Sigma V$ is fixed and $m_\pi L\alt 1$ 
(cf.~table \ref{table:analysis}).
An idea of the size of the simulation can be gotten from the run parameters
listed in Table~\ref{table:run}. 
We note that our method requires very small quark masses and hence many CG
iterations (typically $10^4$ for the smaller mass in $V=6^4$, restarting
every 500) for inverting the fermion matrix; the restriction to small volume,
while allowing easy 
replication of the simulation for many processors, precludes gaining any
advantage from true parallel computation. 

\begin{table*}[htb]
\begin{ruledtabular}
\begin{tabular}{cdddcc}
$V$ & {\it m} &\textrm{trajectories}&dt&\textrm{steps/trajectory}&\textrm{trajectories/measurement}\\\hline
$4^4$&0.01&4000&0.0067&150&1\\
$4^4$&0.025&4000&0.0125&80&1\\
$6^4$&0.002&27960&0.001&100&5\\
$6^4$&0.005&5 135&0.001&200&5
\end{tabular}
\end{ruledtabular}
\caption{\label{table:run}
Run parameters for the Hybrid Monte Carlo simulations.}
\end{table*}

\begin{figure*}[!ht]
\begin{center}
% top figure: m=0.002
\includegraphics*[width=.9\columnwidth]{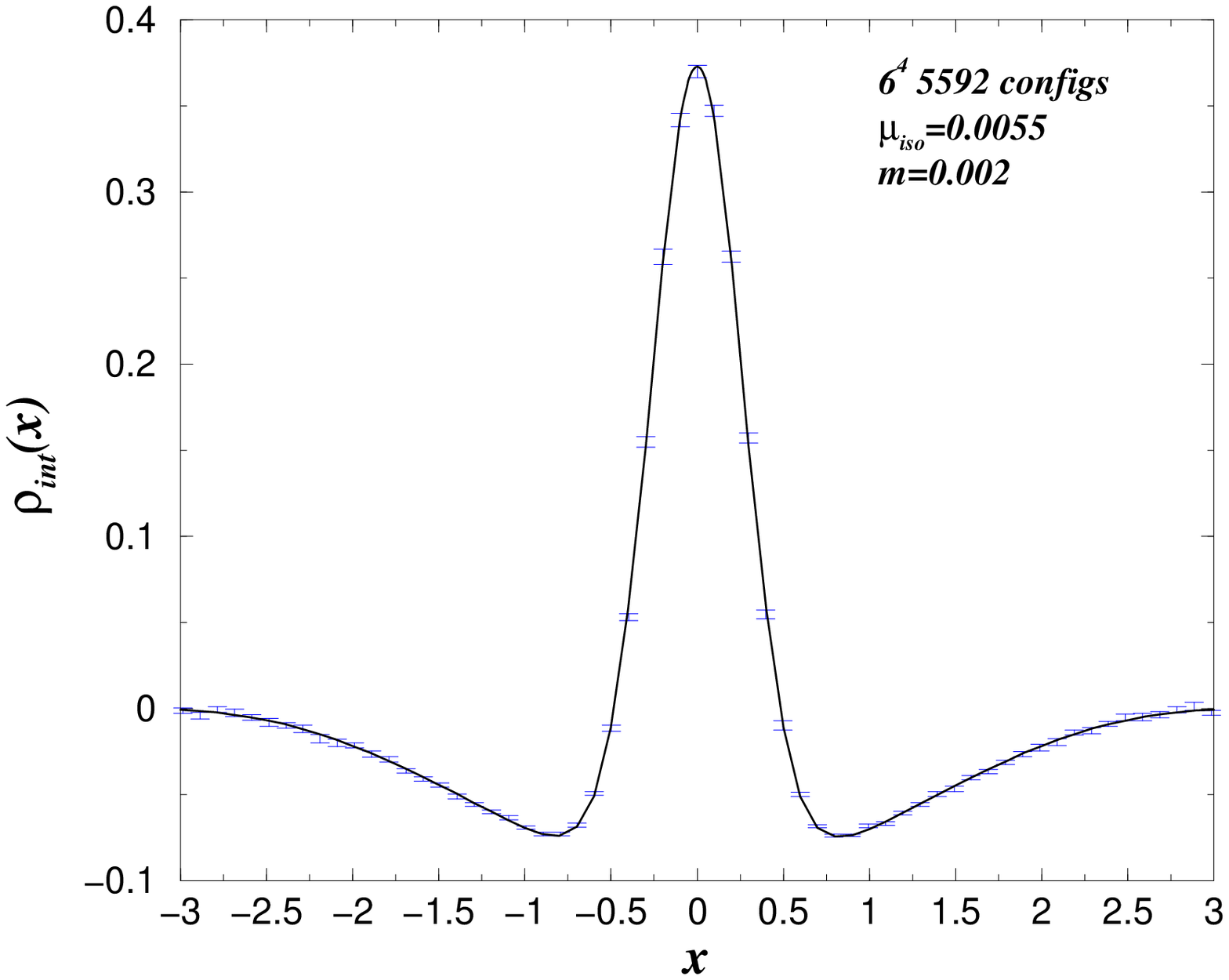}
% bottom figure: m=0.005
\includegraphics*[width=.9\columnwidth]{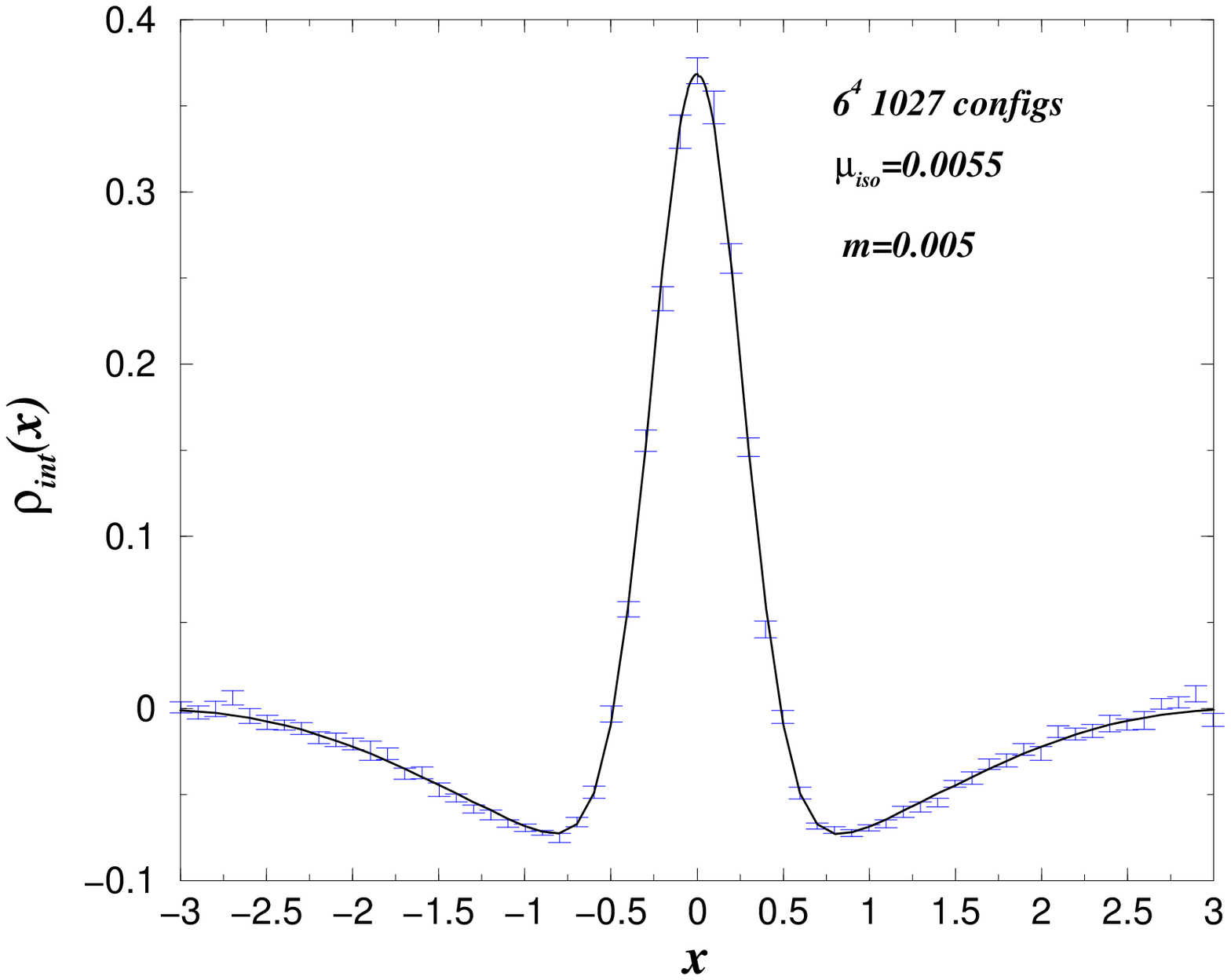}
\caption{\label{fig:intcf6}
  The integrated correlation function (\ref{corr-int})
  with $\txi_{\rm  min}=5$ and $\txi_{\rm max}=55$ for the ensembles with
  $V=6^4$ and $\mui=0.0055$. 
Left: $m=0.002$.  The curve is the result of a fit with
$\Sigma=0.3200\pm0.0005$ and $F_\pi=0.3379\pm0.0018$, with
$\chi^2/\textit{dof}=1.33$.
Right: $m=0.005$. The fit parameters are $\Sigma=0.3212\pm0.0012$ and 
$F_\pi=0.3384\pm0.0025$, with $\chi^2/\textit{dof}=1.13$. }
\end{center}
\end{figure*}

As discussed in \cite{DHSS}, when extracting physical observables we must keep in
mind that the continuum theory describes 4 tastes of quark.
Therefore, to determine the values of $F_\pi$ and $\Sigma$ from the staggered
eigenvalue spectrum we must replace $V$ in the analytical predictions by 4$V$.

We begin with our main numerical result, the pion decay constant $F_\pi$. In
order to improve the statistics we consider the integrated correlation
function
\be\label{corr-int}
\rho^{(2)}_{\rm int}(x)\equiv\int_{\xi_{\rm min}}^{\xi_{\rm max}}{\rm d}\xi\,
\rho^{(2)}_s(\xi+x,\xi,\hat m_u,\hat m_d;i\hat\mui).
\ee
In Fig.~\ref{fig:intcf6} we show this integrated correlation function as
measured on our $6^4$ ensemble, for both masses.
The curves are fits to Eq.~(\ref{corr-int}) via Eq.~(\ref{rho-result}), see
Table \ref{table:analysis}. The weighted averages 
 of the fit parameters for the two data sets are $\Sigma=0.3204\pm0.0004$ and
 $F_\pi=0.3381\pm0.0010$, in lattice units.
There are a number of other ways to determine $\Sigma$ from the eigenvalue
densities and correlations; we will see that the value just obtained is
consistent with other features of the spectra.

\begin{figure*}[!htb]
\begin{center}
% top figure: m=0.002
\includegraphics*[width=.9\columnwidth]{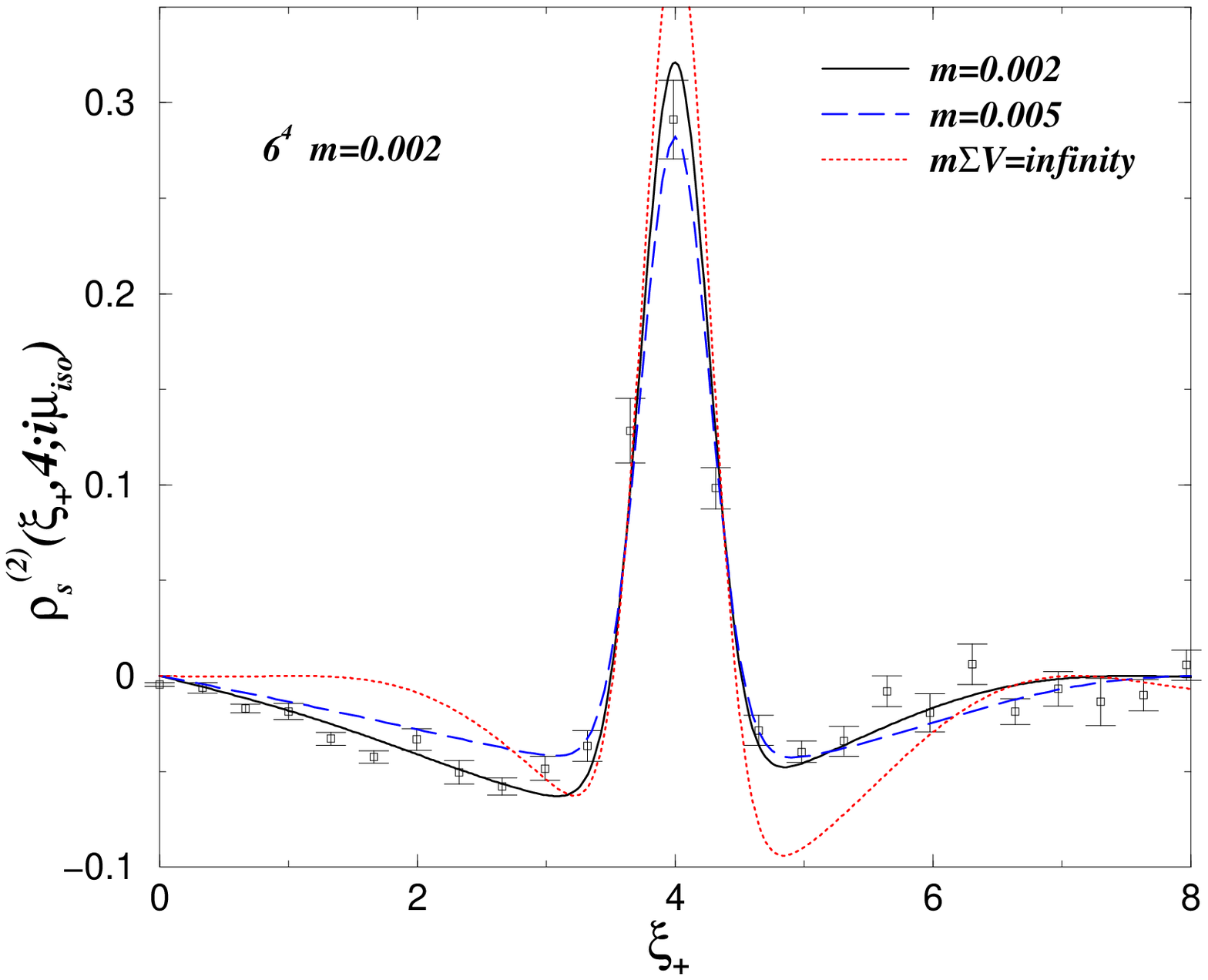}
% bottom figure: m=0.005
\includegraphics*[width=.9\columnwidth]{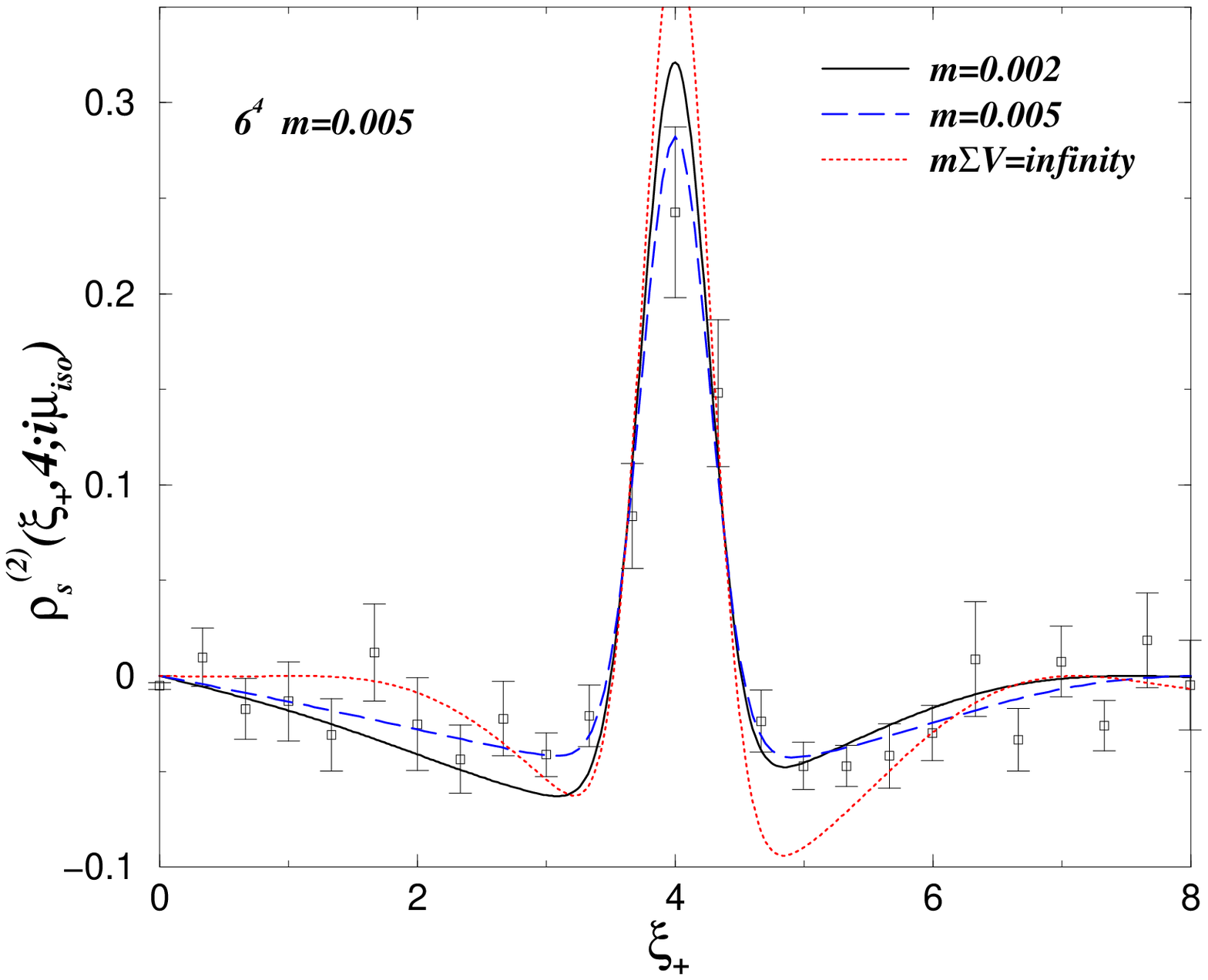}
\caption{\label{fig:cf6}
  The correlation function (\ref{rho2micro}), with $\xi_-$ fixed at 4, measured in the ensembles with $V=6^4$.
Left: $m=0.002$. Right: $m=0.005$. The curves are Eq.~(\ref{rho-result}) with $m=0.002$ (solid), 0.005 (dashed), and~$\infty$ (dotted).}
\end{center}
\end{figure*}

We now proceed to study the sensitivity of the correlation function to the mass.
Since the mass affects mainly low-lying eigenvalues, this sensitivity is
greatest when the correlation function is \textit{not} integrated.
We show in Fig.~\ref{fig:cf6} the measured correlation function with one
eigenvalue fixed at $\xi_-=4$.  For each data set we compare to theoretical
curves [Eq.~(\ref{rho-result})] with masses corresponding to the two chosen
values of $m$ in the simulation, as well as to infinitely high mass (the
quenched theory). We set $\Sigma$ and $F_\pi$ to the values obtained in the
fits shown in Fig.~\ref{fig:intcf6}, for each mass separately.
It is clear that the effect of the mass in the analytical prediction is
consistent with the data and inconsistent with the quenched result. It
would take, however, a somewhat greater number of configurations to get a
precise test of the mass dependence.
\begin{table*}[htb]
\begin{ruledtabular}
\begin{tabular}{cccccc}
$V$ & $m$ & $F_\pi$ & $\Sigma$ &
$m_\pi/F_\pi$ & $m_\pi L$ \\\hline
$6^4$ & 0.002 & $0.3379\pm0.0018$ & $0.3200\pm0.0005$ & $0.3133$ & 0.6354\\
$6^4$ & 0.005 & $0.3384\pm0.0025$ & $0.3212\pm0.0012$ & $0.4949$ & 1.005
\end{tabular}
\end{ruledtabular}
\caption{\label{table:analysis}
Results of the data analysis. To extract the ratio $m_\pi/F_\pi$ we used the
GOR-relation $m_\pi^2F_\pi^2=2m\Sigma$.}
\end{table*} 

\begin{figure*}[!htb]
\begin{center}
% top figures: m=0.01, 0.002
\includegraphics*[width=.9\columnwidth]{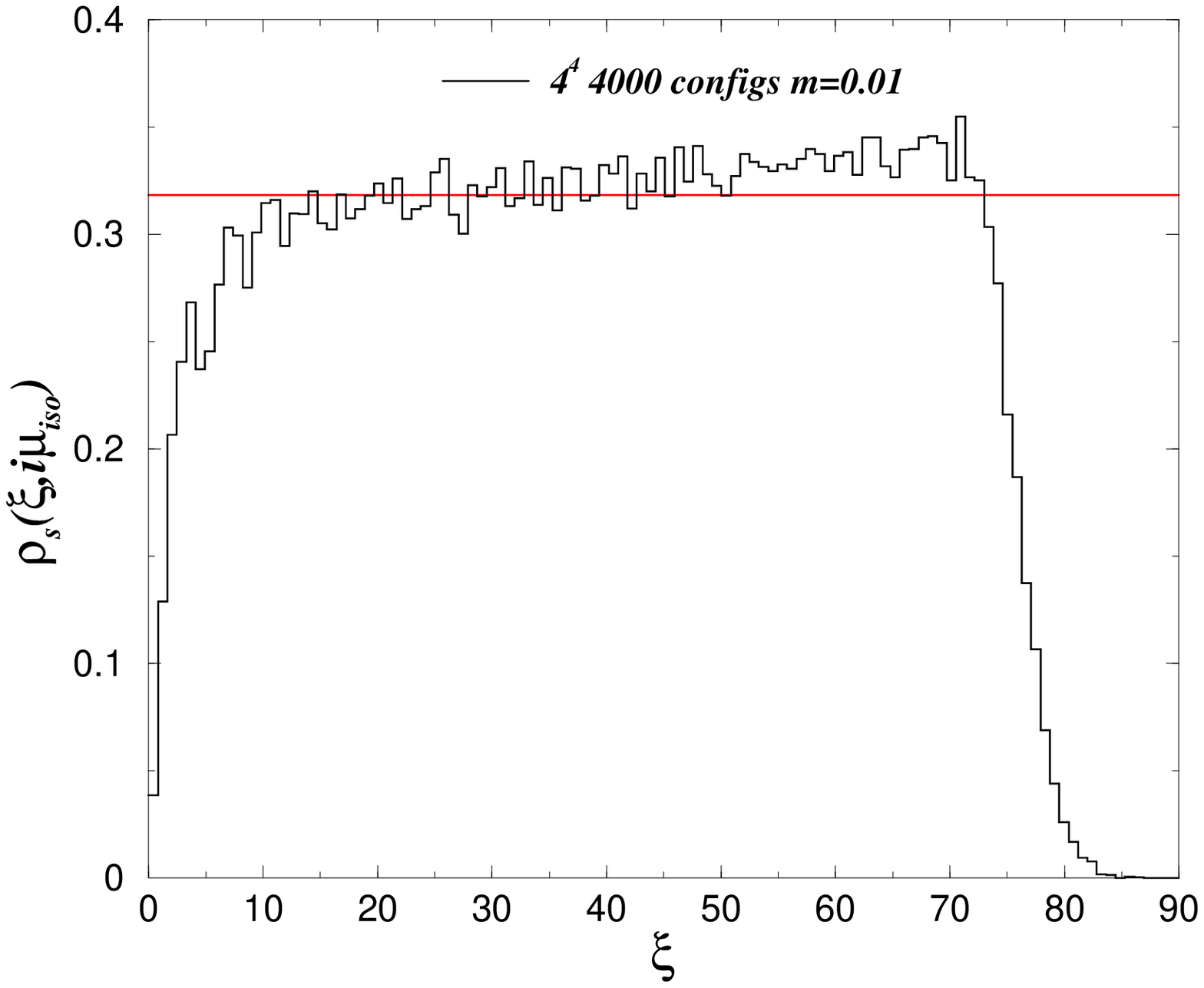}
\includegraphics*[width=.9\columnwidth]{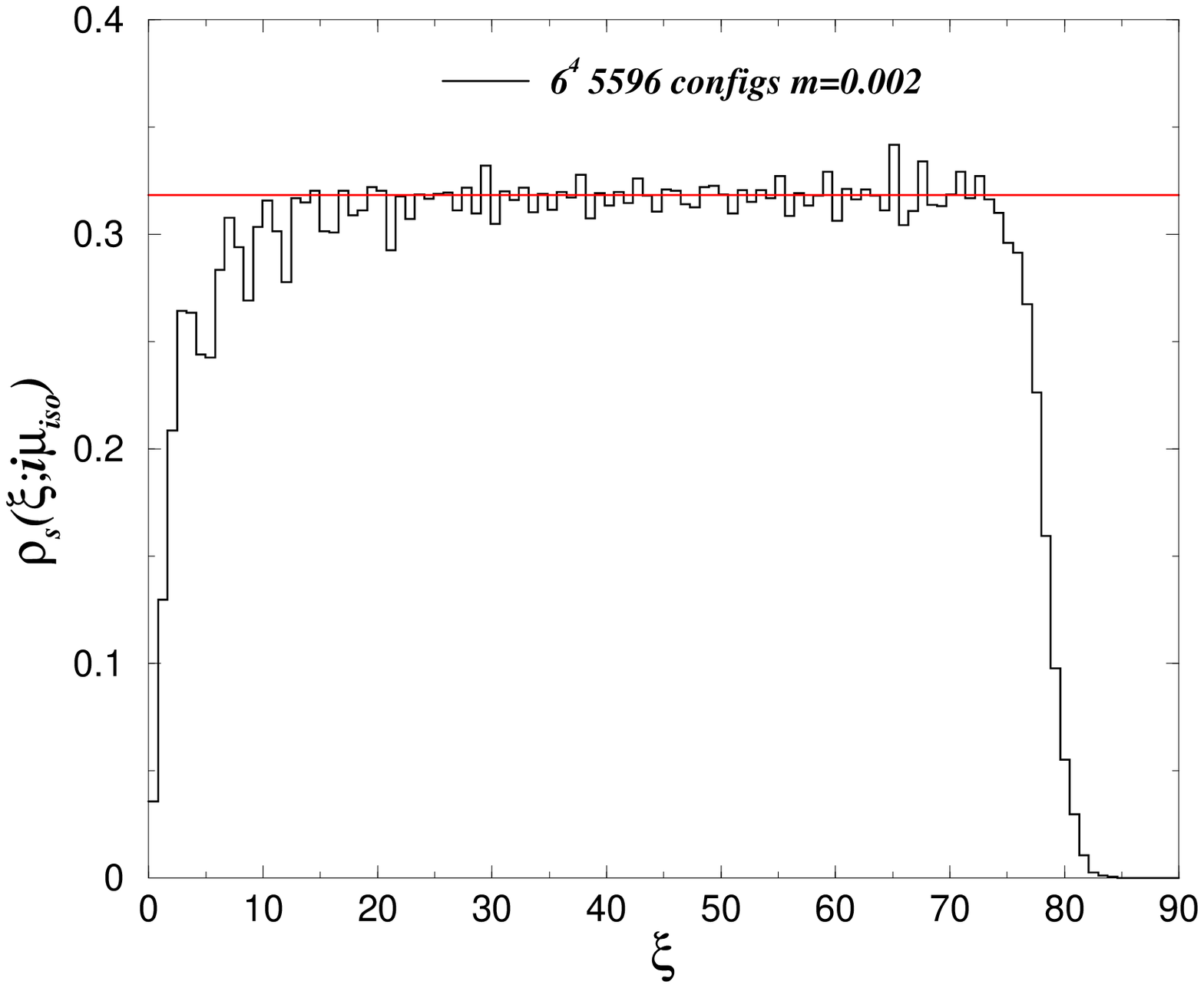}
\caption{\label{fig:dens}
The eigenvalue density of the 24 lowest eigenvalues measured for the smaller
scaled mass: $V=4^4$, $m=0.01$ (left) and $V=6^4$, $m=0.002$ (right).  The
horizontal line in each case is at $\rho_s=1/\pi$. For $V=4^4$ the eigenvalue
density only remains briefly at the height $1/\pi$ whereas for $V=6^4$ all of
the first 24 eigenvalues are within this plateau.}
\end{center}
\end{figure*}

\begin{figure*}[!htb]
\begin{center}
% figures: m=0.01, 0.025
\includegraphics*[width=.9\columnwidth]{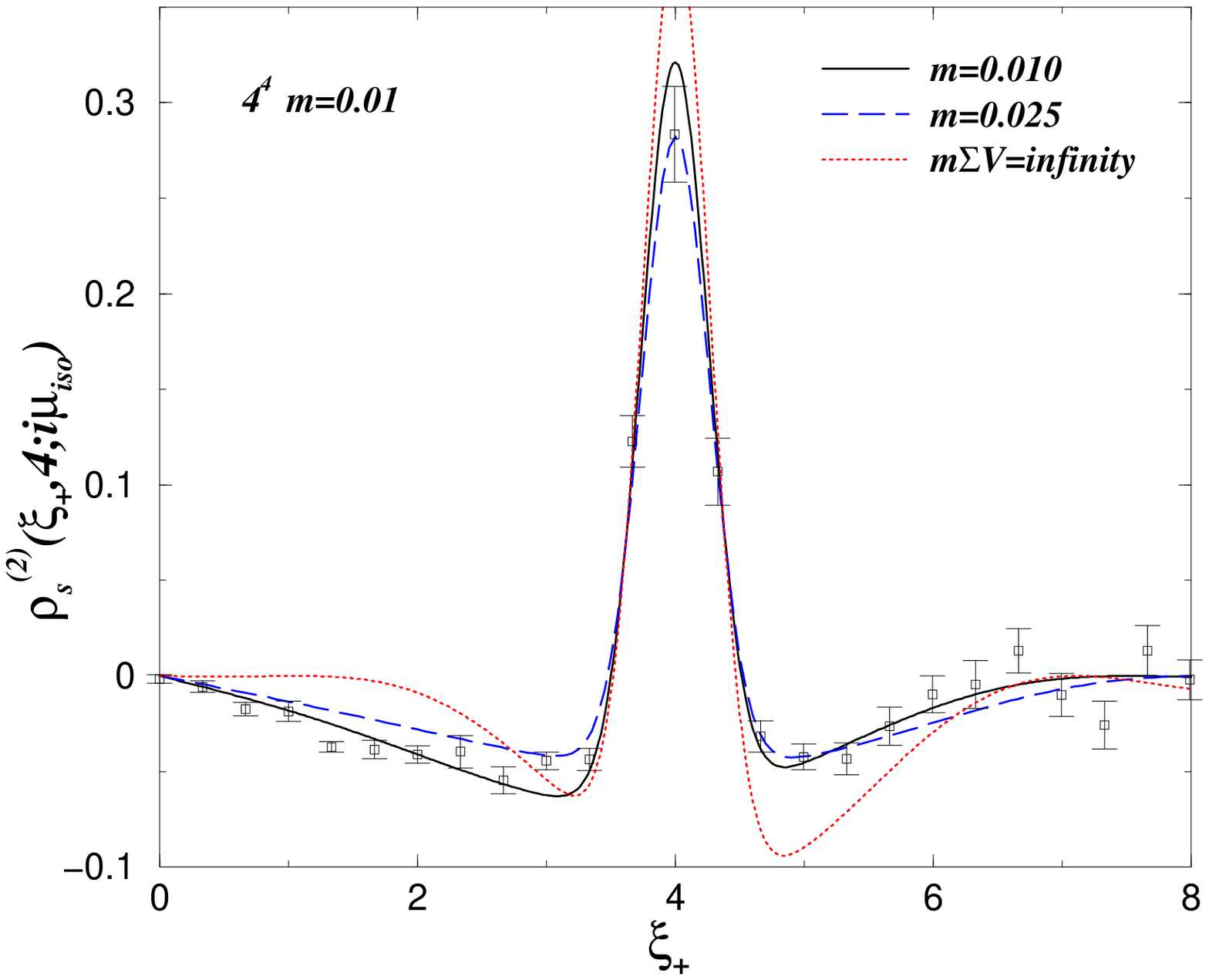}
\includegraphics*[width=.9\columnwidth]{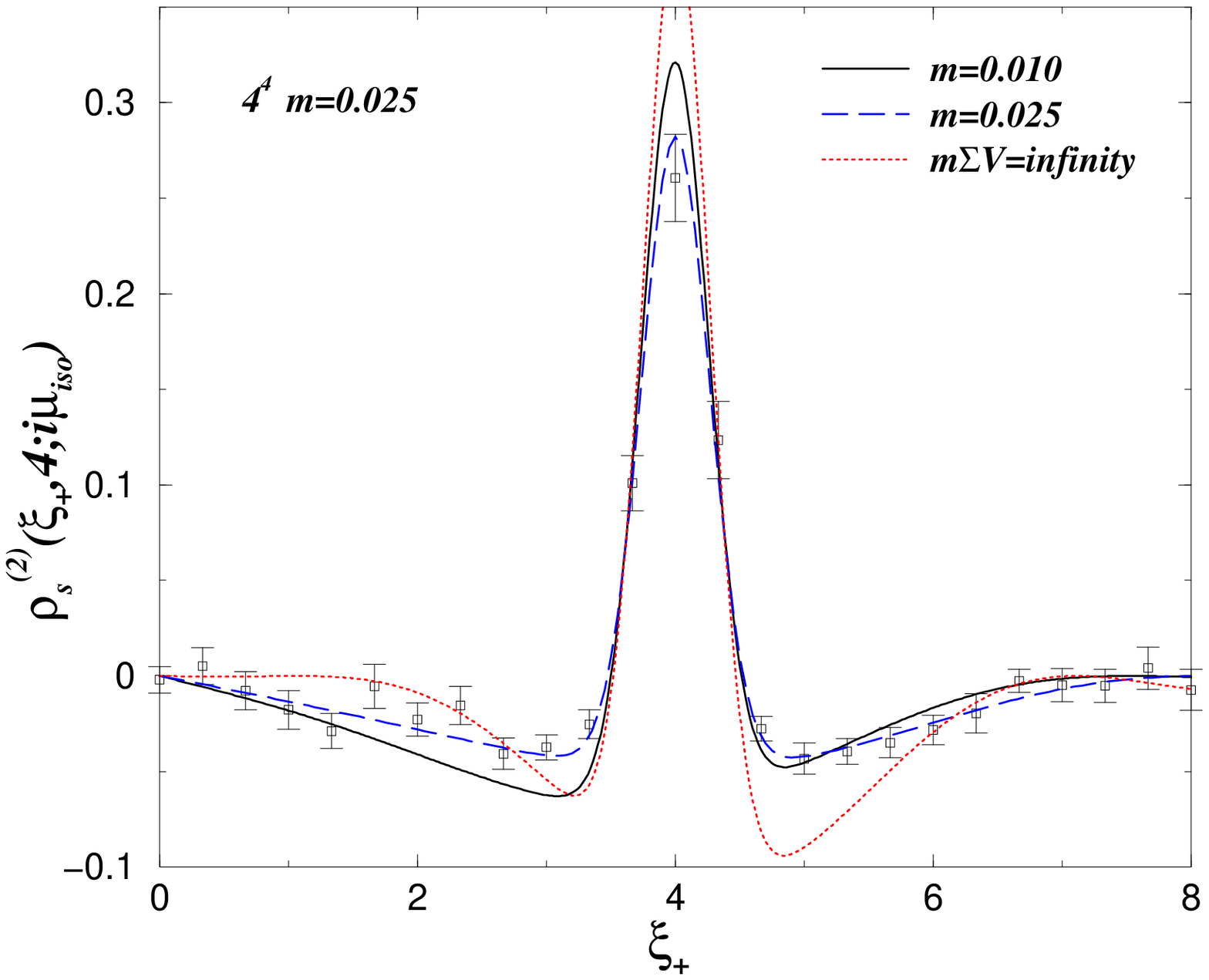}
\caption{\label{fig:corr4i4}
The correlation function (\ref{corr-def}), with $\xi_-$ fixed at 4, for
$V=4^4$; $m=0.01$ (left) to be compared with the full curve and $m=0.025$
(right) to be compared with the dashed curve.  The curves are obtained from 
(\ref{rho-result}) with $\Sigma$  and $F_\pi$ as determined in
Fig.~\ref{fig:intcf6}. The dotted curve indicates the quenched result.}
\end{center}
\end{figure*}

The scaling properties of the $\epsilon$ regime can be studied by comparing distributions for the two volumes.
We begin with the eigenvalue density 
shown in Fig.~\ref{fig:dens}. For the larger volume (right hand figure) the
density as expected rises to a constant level (by convention the value
reached should be $1/\pi$ as indicated by the horizontal line). The fall-off around
the value $\xi=80$ occurs since only the lowest 24 eigenvalues have been
computed. On the other hand the eigenvalue density in the smaller volume (left
hand figure) only briefly remains at the plateau of height $1/\pi$. Thus 
it is not advisable to consider the integrated correlation function over the
entire range of the 24 eigenvalues computed. In Fig.~\ref{fig:corr4i4} 
we consider rather the correlation function with $\xi_-$
fixed at the value 4. Since 
we have chosen the quark masses such that $mV$ is fixed, the
analytical curves shown are identical to those in Fig.~\ref{fig:cf6}. The
general agreement confirms the expected scaling behavior.

\begin{figure}[!htb]
\begin{center}
\includegraphics*[width=.9\columnwidth]{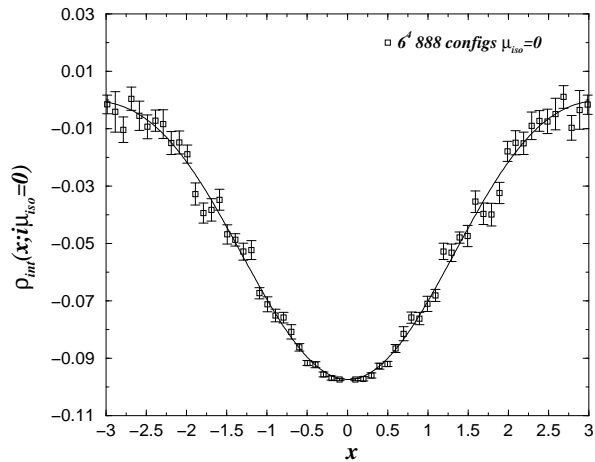}
\caption{\label{fig:mu0}
The integrated correlation function at $\mui=0$, for $V=6^4$; $m=0.002$. The
delta function at $x=0$ is not shown. The comparison with the analytic curve
uses the value of $\Sigma$ determined in Fig.~\ref{fig:intcf6}.}
\end{center}
\end{figure}

Finally we take a brief look at the case where $\mu=0$. In this case the
operators $D_+$ and $D_-$ are identical and the correlation function
consequently has a delta-function peak at equal arguments. In 
Fig.~\ref{fig:mu0} we show
the integrated correlation function at $\mui=0$ as obtained on the $6^4$
lattice at mass $m=0.002$ along with the analytical prediction, using the value of
$\Sigma$ determined in Fig.~\ref{fig:intcf6}. The consistency with the data
is a non-trivial cross check of the value of $\Sigma$ obtained in 
Fig.~\ref{fig:intcf6}.
The delta-function at $\xi_+=\xi_-$ is not shown. As $\mui$ is turned to a
non-zero value the delta function peak gets smeared into a broader peak
around zero as in Fig.~\ref{fig:intcf6}. It is the fit to the shape of
this broader peak that allows us to determine $F_\pi$.

\section{Summary}

We have demonstrated that $F_\pi$ can be extracted with high precision
from small lattices in unquenched lattice QCD. The approach exploits the
dramatic dependence of the universal microscopic correlation functions on
$F_\pi$. This dependence enters through the coupling to an external vector 
field in form of an imaginary isospin chemical potential. The derivation 
of the analytical results used here is
quite involved and will be presented in \cite{nextpaper}.
We have performed appropriate two flavor lattice simulations, 
and the numerical results have been shown to agree with our analytical 
predictions. This agreement has allowed us to extract $F_\pi$ in 
lattice units. See Table \ref{table:analysis} for a summary. 

In this first study of the method with dynamical fermions we have
worked with an unimproved staggered action. It would be of great
interest to perform a simulation with actions suffering less severely
from scaling violations, thus allowing to quote a result for $F_\pi$
in physical units.

\vspace{1cm}

\begin{acknowledgments}
PHD would like to thank the Yukawa
Institute for Theoretical Physics at Kyoto 
University where this work was completed during
the YITP workshop on ``Actions and Symmetries in Lattice Gauge Theory''
YITP-W-05-25. 
The work of KS we financed by the Carlsberg Foundation.
The work of BS was supported in part by the Israel Science
Foundation under grant no.~173/05.  He thanks the Niels Bohr
Institute for its hospitality. 
The work of DT was supported by the NSF under grant
No.~NSF-PHY0304252. He thanks the Particle Physics Group at the
Rensselaer Polytechnic Institute for its hospitality. 
Our computer code is based on the
public lattice gauge theory code of the MILC Collaboration
\cite{MILC1}. 
\end{acknowledgments}


\begin{thebibliography}{0}

\bibitem{LS}J.~Gasser and H.~Leutwyler,
  %``Light Quarks At Low Temperatures,''
  Phys.\ Lett.\ B {\bf 184}, 83 (1987);
  %%CITATION = PHLTA,B184,83;%%
  %``Thermodynamics Of Chiral Symmetry,''
  {\bf 188}, 477 (1987);
  %%CITATION = PHLTA,B188,477;%%
H.~Neuberger,
  %``A Better Way To Measure F(Pi) In The Linear Sigma Model,''
  Phys.\ Rev.\ Lett.\  {\bf 60}, 889 (1988);
  %%CITATION = PRLTA,60,889;%%
H.~Leutwyler and A.~Smilga,
%``Spectrum of Dirac operator and role of winding number in QCD,''
Phys.\ Rev.\ D {\bf 46}, 5607 (1992).
%%CITATION = PHRVA,D46,5607;%%



\bibitem{Jac}E.~V.~Shuryak and J.~J.~M.~Verbaarschot,
  %``Random matrix theory and spectral sum rules for the Dirac operator in
  %QCD,''
  Nucl.\ Phys.\ A {\bf 560}, 306 (1993)
  [hep-th/9212088];
  %%CITATION = HEP-TH 9212088;%%
M.~E.~Berbenni-Bitsch \textit{et al.},
  %``Random-matrix
  %universality in the small-eigenvalue spectrum of the  lattice
  %Dirac operator,''
  Nucl.\ Phys.\ Proc.\ Suppl.\  {\bf 63}, 820 (1998)
  [hep-lat/9709102];
  %%CITATION = HEP-LAT 9709102;%%
P.~H.~Damgaard, U.~M.~Heller and A.~Krasnitz,
  %``Microscopic spectral density of the Dirac operator in quenched QCD,''
  Phys.\ Lett.\ B {\bf 445}, 366 (1999)
  [hep-lat/9810060].
  %%CITATION = HEP-LAT 9810060;%%

\bibitem{Jac1}J.~J.~M.~Verbaarschot,
  %``Universal scaling of the valence quark mass dependence of the chiral
  %condensate,''
  Phys.\ Lett.\ B {\bf 368}, 137 (1996)
  [hep-ph/9509369];
  %%CITATION = HEP-PH 9509369;%%
P.~H.~Damgaard, R.~G.~Edwards, U.~M.~Heller and R.~Narayanan,
  %``Universal scaling of the chiral condensate in finite-volume gauge
  %theories,''
  Phys.\ Rev.\ D {\bf 61}, 094503 (2000)
  [hep-lat/9907016];
  %%CITATION = HEP-LAT 9907016;%%
P.~Hernandez, K.~Jansen and L.~Lellouch,
  %``Finite-size scaling of the quark condensate in quenched lattice QCD,''
  Phys.\ Lett.\ B {\bf 469}, 198 (1999)
  [hep-lat/9907022].
  %%CITATION = HEP-LAT 9907022;%%

\bibitem{DN}
P.~H.~Damgaard and S.~M.~Nishigaki,
  %``Distribution of the k-th smallest Dirac operator eigenvalue,''
  Phys.\ Rev.\ D {\bf 63}, 045012 (2001)
  [hep-th/0006111].
  %%CITATION = HEP-TH 0006111;%%



\bibitem{BB}
  M.~E.~Berbenni-Bitsch, S.~Meyer, A.~Sch\"afer, J.~J.~M.~Verbaarschot and
T.~Wettig,
  %``Microscopic universality in the spectrum of the lattice Dirac
  % operator,''
  Phys.\ Rev.\ Lett.\  {\bf 80} (1998) 1146
  [hep-lat/9704018];
  %%CITATION = HEP-LAT 9704018;%%
R.~G.~Edwards, U.~M.~Heller, J.~E.~Kiskis and R.~Narayanan,
%``Quark spectra, topology and random matrix theory,''
\textit{ibid.}  {\bf 82}, 4188 (1999)
[hep-th/9902117];
%%CITATION = HEP-TH 9902117;%%
R.~G.~Edwards, U.~M.~Heller and R.~Narayanan,
  %``Small eigenvalues of the staggered Dirac operator in the adjoint
  %representation and random matrix theory,''
  Phys.\ Rev.\ D {\bf 60}, 077502 (1999)
  [hep-lat/9902021];
  %%CITATION = HEP-LAT 9902021;%%
P.~H.~Damgaard, U.~M.~Heller, R.~Niclasen and B.~Svetitsky,
  %``Patterns of spontaneous chiral symmetry breaking in vectorlike gauge
  %theories,''
  Nucl.\ Phys.\ B {\bf 633}, 97 (2002)
  [hep-lat/0110028];
  %%CITATION = HEP-LAT 0110028;%%
L.~Giusti, M.~L\"uscher, P.~Weisz and H.~Wittig,
  %``Lattice QCD in the epsilon-regime and random matrix theory,''
  J. High Energy Phys. {\bf 0311}, 023 (2003)
  [hep-lat/0309189].
  %%CITATION = HEP-LAT 0309189;%%

\bibitem{BB1}
M.~E.~Berbenni-Bitsch, S.~Meyer and T.~Wettig,
  %``Microscopic universality with dynamical fermions,''
  Phys.\ Rev.\ D {\bf 58}, 071502 (1998)
  [hep-lat/9804030].
  %%CITATION = HEP-LAT 9804030;%%

\bibitem{DHNR}
P.~H.~Damgaard, U.~M.~Heller, R.~Niclasen and K.~Rummukainen,
  %``Eigenvalue distributions of the QCD Dirac operator,''
  Phys.\ Lett.\ B {\bf 495}, 263 (2000)
  [hep-lat/0007041].
  %%CITATION = HEP-LAT 0007041;%%


\bibitem{Fpi-quen}
%\cite{Aoki:1999av}
%\bibitem{Aoki:1999av}
  S.~Aoki {\it et al.},  %[JLQCD Collaboration],
  %``Pion decay constant for the Kogut-Susskind quark action in quenched
  %lattice QCD,''
  Phys.\ Rev.\ D {\bf 62}, 094501 (2000)
  [hep-lat/9912007];
  %%CITATION = HEP-LAT 9912007;%%
%\cite{Dong:2001fm}
%\bibitem{Dong:2001fm}
  S.~J.~Dong, et al., % T.~Draper, I.~Horvath, F.~X.~Lee, K.~F.~Liu and J.~B.~Zhang,
  %``Chiral properties of pseudoscalar mesons on a quenched 20**4 lattice
  %with
  %overlap fermions,''
  Phys.\ Rev.\ D {\bf 65}, 054507 (2002)
  [hep-lat/0108020];
  %%CITATION = HEP-LAT 0108020;%%
%\cite{Chiu:2003iw}
%\bibitem{Chiu:2003iw}
  T.~W.~Chiu and T.~H.~Hsieh,
  %``Light quark masses, chiral condensate and quark-gluon condensate in
  %quenched lattice QCD with exact chiral symmetry,''
  Nucl.\ Phys.\ B {\bf 673}, 217 (2003)
  [hep-lat/0305016];
  %%CITATION = HEP-LAT 0305016;%%
%\cite{Durr:2005ik}
%\bibitem{Durr:2005ik}
  S.~D\"urr and C.~Hoelbling,
  %``Continuum physics with quenched overlap fermions,''
  Phys.\ Rev.\ D {\bf 72}, 071501 (2005)
  [hep-ph/0508085];
  %%CITATION = HEP-PH 0508085;%%
%\cite{Gattringer:2005ij}
%\bibitem{Gattringer:2005ij}
  C.~Gattringer, P.~Huber and C.~B.~Lang,%  [Bern-Graz-Regensburg (BGR) 
  %          Collaboration],
  %``Lattice calculation of low energy constants with Ginsparg-Wilson type
  %fermions,''
  Phys.\ Rev.\ D {\bf 72}, 094510 (2005)
  [hep-lat/0509003].
  %%CITATION = HEP-LAT 0509003;%%

\bibitem{Fpi-dyn}
%\cite{Allton:2004qq}
%\bibitem{Allton:2004qq}
  C.~R.~Allton {\it et al.},  % [UKQCD Collaboration],
  %``Improved Wilson QCD simulations with light quark masses,''
  Phys.\ Rev.\ D {\bf 70}, 014501 (2004)
  [hep-lat/0403007];
  %%CITATION = HEP-LAT 0403007;%%
%\cite{Farchioni:2004tv}
%\bibitem{Farchioni:2004tv}
  F.~Farchioni, I.~Montvay and E.~Scholz, %  [qq+q Collaboration],
  %``Quark mass dependence of pseudoscalar masses and decay constants on a
  %lattice,''
  Eur.\ Phys.\ J.\ C {\bf 37}, 197 (2004)
  [hep-lat/0403014];
  %%CITATION = HEP-LAT 0403014;%%
%\cite{Aubin:2004fs}
%\bibitem{Aubin:2004fs}
  C.~Aubin {\it et al.}, %  [MILC Collaboration],
  %``Light pseudoscalar decay constants, quark masses, and low energy
  %constants
  %from three-flavor lattice QCD,''
  Phys.\ Rev.\ D {\bf 70}, 114501 (2004)
  [hep-lat/0407028];
  %%CITATION = HEP-LAT 0407028;%%
%\cite{Lin:2005gh}
%\bibitem{Lin:2005gh}
  M.~F.~Lin,
  %``Probing the chiral limit of M(pi) and f(pi) in 2+1 flavor QCD with domain
  %wall fermions from QCDOC,''
  PoS {\bf LAT2005}, 094 (2005)
  [hep-lat/0509178].
  %%CITATION = HEP-LAT 0509178;%%


\bibitem{DHSS}
  P.~H.~Damgaard, U.~M.~Heller, K.~Splittorff and B.~Svetitsky,
  %``A new method for determining F(pi) on the lattice,''
  Phys.\ Rev.\ D {\bf 72}, 091501 (2005)
  [hep-lat/0508029].
  %%CITATION = HEP-LAT 0508029;%%


\bibitem{MT}
T.~Mehen and B.~C.~Tiburzi,
  %``Quarks with twisted boundary conditions in the epsilon regime,''
  Phys.\ Rev.\ D {\bf 72}, 014501 (2005)
  [hep-lat/0505014].
  %%CITATION = HEP-LAT 0505014;%%

\bibitem{OW}
J.~C.~Osborn and T.~Wettig,
  %``Dirac eigenvalue correlations in quenched QCD at finite density,''
  Proc. Sci. {\bf LAT2005}, 200 (2005)
  [hep-lat/0510115].
  %%CITATION = HEP-LAT 0510115;%%


\bibitem{baryon}
G. Akemann and G. Vernizzi, Nucl. Phys. B { 660} (2003) 532 [hep-th/0212051];
J. C. Osborn, Phys. Rev. Lett. {\bf 93}, 222001 (2004)
[hep-th/0403131];
%%CITATION = HEP-TH 0403131;%%
G.~Akemann, J.~C.~Osborn, K.~Splittorff and J.~J.~M.~Verbaarschot,
%``Unquenched QCD Dirac operator spectra at nonzero baryon chemical
%potential,''
Nucl.\ Phys. \ B {\bf 712} (2005) 287 
[hep-th/0411030].
%%CITATION = HEP-TH 0411030;%%.


\bibitem{SplitVerb2} K. Splittorff and J.J.M. Verbaarschot,
 Nucl. Phys. B {\bf 683}, 467 (2004),
[hep-th/0310271].
%%CITATION = HEP-TH 0310271;%%

\bibitem{Fpi-ep}
W.~Bietenholz et al.,
  %``Axial correlation functions in the epsilon-regime: A numerical study  with
  %overlap fermions,''
  JHEP {\bf 0402}, 023 (2004)
  [hep-lat/0311012];
  %%CITATION = HEP-LAT 0311012;%%
L.~Giusti, P.~Hernandez, M.~Laine, P.~Weisz and H.~Wittig,
  %``Low-energy couplings of QCD from current correlators near the chiral
  %limit,''
  JHEP {\bf 0404}, 013 (2004);
  [hep-lat/0402002].
  %%CITATION = HEP-LAT 0402002;%%
H.~Fukaya, S.~Hashimoto and K.~Ogawa,
  %``Low-lying mode contribution to the quenched meson correlators in the
  %epsilon-regime,''
  Prog.\ Theor.\ Phys.\  {\bf 114}, 451 (2005)
  [hep-lat/0504018];
  %%CITATION = HEP-LAT 0504018;%%
S.~Shcheredin and W.~Bietenholz,
  %``Low energy constants from the zero mode contribution to the pseudo-scalar
  %correlator,''
  PoS {\bf LAT2005}, 134 (2005)
  [hep-lat/0508034].
  %%CITATION = HEP-LAT 0508034;%%


\bibitem{DOTV}
P.~H.~Damgaard, J.~C.~Osborn, D.~Toublan and J.~J.~M.~Verbaarschot,
  %``The microscopic spectral density of the {QCD} Dirac operator,''
  Nucl.\ Phys.\ B {\bf 547}, 305 (1999)
  [hep-th/9811212];
  %%CITATION = HEP-TH 9811212;%%
D.~Toublan and J.~J.~M.~Verbaarschot,
%``Statistical properties of the spectrum of the QCD Dirac operator at low
%energy,''
\textit{ibid.} {\bf 603} (2001) 343
[hep-th/0012144].
%%CITATION = HEP-TH 0012144;%%

%UMH: reordered according to order of citation in text
\bibitem{DS}
P.~H.~Damgaard and K.~Splittorff,
%``Partially quenched chiral perturbation theory and the replica method,''
Phys.\ Rev.\ D {\bf 62}, 054509 (2000)
[hep-lat/0003017].
%%CITATION = HEP-LAT 0003017;%%

\bibitem{K}E.~Kanzieper,
  %``Replica field theories, Painleve transcendents and exact correlation
  %functions,''
  Phys.\ Rev.\ Lett.\  {\bf 89}, 250201 (2002)
  [cond-mat/0207745].
  %%CITATION = COND-MAT 0207745;%%

\bibitem{SplitVerb1} K. Splittorff and J.J.M. Verbaarschot,
  Phys. Rev. Lett. {\bf 90}, 041601 (2003)
[cond-mat/0209594];
%%CITATION = COND-MAT 0209594;%%
Nucl.\ Phys.\ B {\bf 695}, 84 (2004)
  [hep-th/0402177].
  %%CITATION = HEP-TH 0402177;%%

\bibitem{AFV}
  G.~Akemann, Y.~V.~Fyodorov and G.~Vernizzi,
  %``On matrix model partition functions for QCD with chemical potential,''
  Nucl.\ Phys.\ B {\bf 694} (2004) 59
  [hep-th/0404063].
  %%CITATION = HEP-TH 0404063;%%

\bibitem{nextpaper}
P.~H.~Damgaard, U.~M.~Heller, K.~Splittorff, B.~Svetitsky, and D. Toublan, in
preparation. 


\bibitem{8flavors}  J.~B.~Kogut and D.~K.~Sinclair,
  %``SU(2) And SU(3) Lattice Gauge Theories With Many Fermions,''
  Nucl.\ Phys.\ B {\bf 295}, 465 (1988);
  %%CITATION = NUPHA,B295,465;%%
  F.~R.~Brown, {\em et al.},
  %``Lattice QCD with eight light quark flavors,''
  Phys.\ Rev.\ D {\bf 46}, 5655 (1992)
  [hep-lat/9206001].
  %%CITATION = HEP-LAT 9206001;%%

\bibitem{DHNR_topo}
P.~H.~Damgaard, U.~M.~Heller, R.~Niclasen and K.~Rummukainen,
  %``Staggered Fermions and Gauge Field Topology,''
  Phys.\ Rev.\ D {\bf 61}, 014501 (2000)
  [hep-lat/9907019].
  %%CITATION = HEP-LAT 9907019;%%

\bibitem{HMC}
  S.~Duane, A.~D.~Kennedy, B.~J.~Pendleton and D.~Roweth,
  %``Hybrid Monte Carlo,''
  Phys.\ Lett.\ B {\bf 195}, 216 (1987);
  %%CITATION = PHLTA,B195,216;%%
  R.~Gupta, G.~W.~Kilcup and S.~R.~Sharpe,
  %``Tuning The Hybrid Monte Carlo Algorithm,''
  Phys.\ Rev.\ D {\bf 38}, 1278 (1988).
  %%CITATION = PHRVA,D38,1278;%%


\bibitem{MILC1}
Available from\\ {\tt http://www.physics.utah.edu/$^\sim$detar/milc/}


\end{thebibliography}
\end{document}